\documentclass[journal]{IEEEtran}
 
\usepackage{xcolor,soul,framed} 
\usepackage{verbatim} 

\colorlet{shadecolor}{yellow}

\ifCLASSINFOpdf
\else
   \usepackage[dvips]{graphicx}
\fi
\usepackage{url}
\usepackage{graphicx}

\graphicspath{{../pdf/}{../jpeg/}}
\DeclareGraphicsExtensions{.pdf,.jpeg,.png}

\usepackage[cmex10]{amsmath}

\usepackage{array}
\usepackage{mdwmath}
\usepackage{mdwtab}
\usepackage{eqparbox}
\usepackage{url}
\usepackage{multirow}
\usepackage{wrapfig}
\usepackage[center]{caption}
\usepackage{caption}
\usepackage{subcaption}
\usepackage{cite}
\usepackage{amsmath}
\usepackage{tikz}
\usetikzlibrary{positioning}
\usepackage{float}

\usepackage{amsmath, amssymb, latexsym}

\usepackage{xcolor}
\definecolor{fc}{HTML}{1E90FF}
\definecolor{h}{HTML}{228B22}
\definecolor{bias}{HTML}{87CEFA}
\definecolor{noise}{HTML}{8B008B}
\definecolor{conv}{HTML}{FFFF99}
\definecolor{pool}{HTML}{B22222}
\definecolor{down}{HTML}{CCFFCC}
\definecolor{view}{HTML}{FFFFFF}
\definecolor{Tconv}{HTML}{FFB266}
\tikzset{fc/.style={black,draw=black,fill=fc,rectangle,minimum height=1cm}}
\tikzset{h/.style={black,draw=black,fill=h,rectangle,minimum height=1cm}}
\tikzset{bias/.style={black,draw=black,fill=bias,rectangle,minimum height=1cm}}
\tikzset{noise/.style={black,draw=black,fill=noise,rectangle,minimum height=1cm}}
\tikzset{Tconv/.style={black,draw=black,fill=Tconv,rectangle,minimum height=0.75cm}}
\tikzset{pool/.style={black,draw=black,fill=pool,rectangle,minimum height=1cm}}
\tikzset{down/.style={black,draw=black,fill=down,rectangle,minimum height=0.75cm}}
\tikzset{view/.style={black,draw=black,fill=view,rectangle,minimum height=.75cm}}
\tikzset{conv/.style={black,draw=black,fill=conv,rectangle,minimum height=.75cm}}
\usetikzlibrary{arrows}
 
\usepackage{xspace}

\begin{document}
\bstctlcite{IEEEexample:BSTcontrol}
    \title{Learning-based Noise Component Map Estimation for Image Denoising}
  \author{Sheyda~Ghanbaralizadeh Bahnemiri,
      Mykola Ponomarenko,\\
      and Karen Egiazarian,~\IEEEmembership{Fellow,~IEEE}

  \thanks{Authors are with Tampere University, Finland,
(e-mail: sheyda.ghanbaralizadehbahnemiri@tuni.fi; mykola.ponomarenko@tuni.fi;
karen.eguiazarian@tuni.fi).  }

  }


\maketitle

\begin{abstract}
A problem of image denoising when images are corrupted by a non-stationary noise is considered in this paper. 
Since in practice no a priori information on noise is available,  noise statistics should be pre-estimated for image denoising. In this paper,  deep convolutional neural network (CNN) based method for  estimation of a map of local, patch-wise, standard deviations of noise (so-called \textit{sigma-map}) is proposed. It achieves the  state-of-the-art performance in accuracy of estimation of  sigma-map for the case of  non-stationary noise, as well as estimation of noise variance for the case of additive white Gaussian noise.  
Extensive experiments on image denoising using estimated sigma-maps  demonstrate that our method outperforms recent CNN-based blind image denoising methods by up to 6 dB in PSNR, as well as other state-of-the-art methods based on sigma-map estimation by up to 0.5 dB, providing same time better usage flexibility. Comparison with the ideal case, when denoising is applied using ground-truth  sigma-map, shows that a difference of corresponding PSNR values for most of noise levels is within 0.1-0.2 dB and does not exceeds 0.6 dB.

\end{abstract}


\IEEEpeerreviewmaketitle

\section{Introduction}

\IEEEPARstart{I}{mage} denoising is one of the most studied   problems of image processing.  Acquired images are often exposed to noise due to low light intensity, camera quality, transmission errors,
etc. \cite{astola2020fundamentals}. 
 During the last decades, a large number of denoising methods have been proposed, such as methods based on 
 local transform domain (e.g. sliding Discrete Cosine Transform (DCT) filtering)  \cite{oktem2007image ,pogrebnyak2012wiener ,foi2007pointwise} and wavelet-based   \cite{oktem1998signal, donoho1998minimax,portilla2003image}) methods,  nonlocal collaborative denoisers   (e.g. BM3D  \cite{dabov2007image})\cite{buades2004image,mahmoudi2005fast}, and learning  (e.g.  Convolutional Neural Networks (CNN)) based  methods \cite{jain2008natural,zhang2017beyond ,zhang2018ffdnet ,guo2019toward,chen2018image, isogawa2017deep, wang2021channel}.
Most of these  denoisers  assume that noise is Additive White Gaussian Noise (AWGN), and  its standard deviation $\sigma$ 
is either known or pre-estimated ($\sigma$ is a constant for the whole image)
\cite{danielyan2009noise,liu2012noise,pyatykh2012image,ponomarenko2018blind}. 

In this paper, we assume that the observed image is corrupted by a noise with  non-i.i.d. pixel-wise Gaussian distribution (similar to the noise model  considered in  \cite{yue2019variational}): 
\begin{equation} \label{eq1}
  y_{i,j} \sim  \mathcal{N}(y_{i,j}|x_{i,j}, \sigma_{i,j}^{2}) ,  i=1,2,...,L, j=1,2,...,K.
\end{equation}
 where \(\mathcal{N}(\cdot | \mu , \sigma^{2})\), here and later in this paper,   denotes the Gaussian distribution with mean \(\mu\) and variance \(\sigma^{2}\); $x_{i,j}$ and $y_{i,j}$ are $(i,j)^{th}$ pixel values of unknown clean image $\mathbf{X}$ and noisy image $\mathbf{y}$, respectively; $L \times K$ is image size, and a  matrix $ \mathbf{\Sigma }$ $=  [ \sigma_{i,j}] $ of size $L \times K$ defines the so-called {\textit{sigma-map}} \cite{yue2019variational}. Note, that in the case of i.i.d. AWGN, sigma-map  becomes a  matrix with all elements having value $\sigma$ (noise standard deviation). \\
 The main goal of the paper is to estimate a sigma-map from a given noisy image, with a further  use of it as an auxiliary  input in image denoising methods.
A special attention will be given to the case when a noise level is  small, as the  most  appealing   from the practical point of view as well as most  difficult case for existing sigma-map estimation methods as it will be demonstrated later on.
\\
Early methods of sigma-map estimation perform  
robust analysis of  transform coefficients of image patches    \cite{lukin2010discrete}, \cite{shulev2006threshold}. However, these  methods often produce large estimation errors for fine details, edges and textures, and, thus, they are not applicable to the case when an image is corrupted by a small level of  noise. \\
During the last decade, CNN becomes a wide-spread tool in image processing and analysis. In  \cite{yue2019variational}, neural  networks (VDNet) for sigma-map estimation and noise suppression were trained jointly, but estimated sigma-map can be used also by itself to be used with other image denoising methods.  VDNet   demonstrates the state-of-the-art performance for sigma-map estimation.
A deep convolutional autoencoder, DCAE, for sigma-map estimation was proposed in \cite{bahncmiri2019deep}. Advantage of this method is a small network size (only 5 Mb), accompanied by a good estimation accuracy.\\
One popular trend in image denoising is to develop fully blind methods which do not require any input  noise parameters \cite{zhang2017beyond}, \cite{guo2019toward}. However, these  methods are not able to provide a quality of denoising comparable to those of the  state-of-the-art methods which use  sigma-map as an auxiliary input. \\ 
 In this paper, we propose a CNN-based method for sigma-map estimation. It consists of the combination of  U-Net \cite{ronneberger2015u} and ResNet \cite{he2016deep}  as a network architecture, an innovative   strategy to prepare and collect images for the training set, and an improved method of patches generation in the custom training loop. It allows to achieve a 
 superior accuracy of sigma-map estimation, outperforming state-of-the-art methods. \\
A structure of the paper is as follows. The proposed sigma-map estimator, a preparation of the training test set and the training process are described in Section II. In Sections III and IV we analyse the results of  sigma-map estimation and its use in image denoising. Section V presents the conclusion of the paper.  

\begin{figure*}[h]
  \centering
  \hspace*{-0.75cm}
  \begin{tikzpicture}
    \node (y) at (0,0) {\small$y$};
    \node[view,rotate=90,minimum width=3.5cm] (inp) at (-1.25,0) {\small$\text{Input noisy image } \mathbf{Y}$};
    
    \node[down,rotate=90,minimum width=3.5cm] (d1) at (0,0) {\small$\text{Cascade of residual blocks}$};
    \node[conv,rotate=90,minimum width=3.5cm] (conv1) at (1,0) {\small$\text{SConv}$};
    
    \node[down,rotate=90,minimum width=3cm] (d2) at (2.25,0){}; 
    \node[conv,rotate=90,minimum width=3cm] (conv2) at (3.25,0){}; 
    
    \node[down,rotate=90,minimum width=2.5cm] (d3) at (4.5,0){};
    \node[conv,rotate=90,minimum width=2.5cm] (conv3) at (5.5,0){};
    
    \node[down,rotate=90,minimum width=2 cm] (d4) at (6.75,0){};
    \node[Tconv,rotate=90,minimum width=2cm] (conv4) at (7.75,0){\small$\text{Tconv}$};%
    
     \node[down,rotate=90,minimum width=2.5 cm] (d5) at (9,0){};
    \node[Tconv,rotate=90,minimum width=2.5cm] (conv5) at (10,0){};
    
     \node[down,rotate=90,minimum width=3cm] (d6) at (11.25,0){};
    \node[Tconv,rotate=90,minimum width=3cm] (conv6) at (12.25,0){};
    
    \node[down,rotate=90,minimum width=3.5 cm] (d7) at (13.5,0){};
    
    \node[view,rotate=90,minimum width=3.5cm] (out) at (14.75,0) {\small$\text{Sigma-map }\mathbf{M_{e}}$};

    \draw[thick , -latex] (inp) -- (d1);
    
    \draw[thick ,-latex] (d1) -- (conv1);
    \draw[thick ,-latex] (conv1) -- (d2);
    
    \draw[thick ,-latex] (d2) -- (conv2);
    \draw[thick ,-latex] (conv2) -- (d3);
    
    \draw[thick ,-latex] (d3) -- (conv3);
    \draw[thick ,-latex] (conv3) -- (d4);

    \draw[thick ,-latex] (d4) -- (conv4);
    \draw[thick ,-latex] (conv4) -- (d5);
    
    \draw[thick ,-latex] (d5) -- (conv5);
    \draw[thick ,-latex] (conv5) -- (d6);
    
    \draw[thick ,-latex] (d6) -- (conv6);
    \draw[thick ,-latex] (conv6) -- (d7);
    \draw[thick ,-latex] (d7) -- (out);

   \draw[thick ,-latex] (-.7, 0)  -- (-.7, 2.5) -- (14.1,2.5)  -- (14.1,0) ;
   \draw[thick ,-latex] (1.6, 0)  -- (1.6, 2)  --node[above]{skip connections} (12.8,2)  -- (12.8,0) ; 
   \draw[thick ,-latex] (3.8, 0)  -- (3.8, 1.6) -- (10.55,1.6)  -- (10.55,0) ;
   \draw[thick ,-latex] (6.1, 0)  -- (6.1, 1.3) -- (8.3,1.3)  -- (8.3,0) ;
   \draw[thick, dashed,-latex] (0, -2)  -- (6.8, -1.5) node[draw=none, midway, below=10pt]{Downscaling}; 
   \draw[thick, dashed,-latex] (7.5, -1.5)  -- (13.5, -2)  node[draw=none, midway, below=9pt]{Upscaling} ;

  \end{tikzpicture}
  \vskip 6px
  \caption{Structural scheme of proposed SDNet (SConv - stride convolution , TConv - transpose convolution)}
  \label{fig:snet}
\end{figure*}
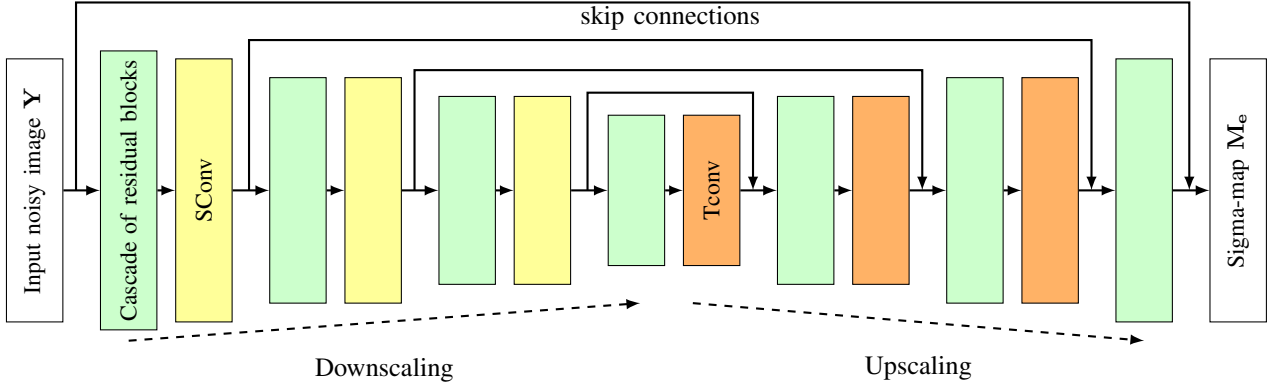


\section{Proposed network for blind sigma-map estimation}

\subsection{Network Design}
A structural scheme of the proposed neural network, SDNet, is presented in Fig.\ref{fig:snet}. 
Inspired by the  DRUNet \cite{zhang2020plug} network architecture,
SDNet combines U-Net \cite{ronneberger2015u} and ResNet \cite{he2016deep} in a single design. SDNet suits well to our task of sigma-map estimation since  analysis is performed at  several image scales. In the middle part of SDNet, each pixel of activations corresponds to 8x8 pixels of the input image. This part of SDNet provides good estimation accuracy for large homogeneous areas. At the same time, a pixel-wise analysis in the last quarter of SDNet provides good robustness to fine details and textures. 

A detailed structure of SDNet is available 
in GitHub page \url{https://github.com/SheydaGhb/SDNet}. SDNet is trained there for the input size of 128\(\times\)128 but it can work also with images of any size. 
 
\subsection{Image Set for Training}
To train SDNet, we  have generated and accurately selected 4238 images from  different datasets as it is described below.

The first 1000 images are selected from Tampere21 database  \cite{ponomarenko2018blind}, which contains 
color images of sizes 1200x800, 960x640 and 768x512. These images are  generated from initial high quality images captured by Canon EOS 250D camera. 
Depending on ISO value, a central part of each captured image was cropped and downscaled (with the downsampling ratio of 5, 6 or 7 times, by averaging pixel values for the regions of 5x5 pixels, 6x6 pixels or 7x7 pixels, respectively). Thus, the resulting images are near noise-free since their  noise variances are  25, 36 or 49 times, respectively,  smaller than those of the  captured images.

The second set of 1000 images is selected from four datasets: KonIQ10k \cite {hosu2020koniq}, FLIVE \cite{bovik2020flive}, NRTID \cite{ponomarenko2010statistical} and SPAQ \cite{fang2020perceptual}. The  merged mean opinion scores (MOS) \cite{kaipio2020merging} of these four databases has been used to select high quality images having the best MOS values. 

The remaining images are collected from the following sources: 
217 images from Flickr2K database \cite{lim2017enhanced}, 123 images from Waterloo Exploration Database \cite{ma2017waterloo}, 103 images from DIV2k database \cite{agustsson2017ntire}, and 1795 images from different photo hostings. 
We have used  no-reference image  quality metric KonCept512 \cite {hosu2020koniq} pre-trained on six databases (KonIQ10k \cite {hosu2020koniq}, Live-in-the-Wild \cite{ghadiyaram2015massive}, FLIVE \cite{bovik2020flive}, NRTID \cite{ponomarenko2010statistical}, HTID \cite{ponomarenko2021color} and SPAQ \cite{fang2020perceptual}) to collect high quality images from the above mentioned image sources. All collected images are of high quality (almost  noise-free) due to good  sensitivity of the pre-trained KonCept512 metric to the presence of noise.

\subsection{Custom Training Loop}
In this paper a  relative error 
$\varepsilon_{m}$ is used as a  qualitative criterion of estimated sigma-map: 

\begin{equation} \label{eq5}
   \varepsilon_{m} = \dfrac{||\pmb{M_{e}} - \pmb{M_{t}}||_{2}}{n||\pmb{M_{t}}||_{2}}
\end{equation}
where $\pmb{M_{e}}$ and $\pmb{M_{t}}$ are the estimated maps and the ground truth map, respectively, $n$ is a number of images in the test set. 
To achieve good quality of image  denoising with the  pre-estimated sigma-map,  
a relative error $\varepsilon_{m}$ has to be smaller than a pre-defined  threshold value (this value was found empirically to be 0.1  \cite{lukin2011methods}). 
This condition cannot be guaranteed if one will use the  conventional  custom loss functions, since the pre-trained network may produce too large relative errors for small noise standard deviations. 
To overcome this, we set the mean variance of sigma-map,  $\sigma_{av}^{2}$ for generated patches to follow a half-normal  distribution: 
\begin{equation} \label{sigm1}
\sigma_{av}^{2} = |\mathcal{N}(0,R^{2})|, 
\end{equation}
where we set $R=40$ for the training.

This provides a larger number of small $\sigma_{av}$ values, and, therefore, their larger weights in the training. As a result, SDNet will better minimize a  prediction error for small $\sigma_{av}$ values, while providing a  comparable $\varepsilon_{m}$ for large $\sigma_{av}$. At the same time, the pre-trained SDNet should give 
good prediction results also for very large $\sigma_{av}$ values (over 100).  
This shall empirically prove an efficiency of our training strategy.

Authors of  \cite{yue2019variational} have used smooth 
surfaces to generate  $\pmb{M_{t}}$ values. However, in practice, this may limit an  applicability of the pre-trained network. In this paper, we have generated the  $\pmb{M_{t}}$ for an input patch by: 
\begin{equation} \label{sigm2}
   \sigma_{i,j} = (\sigma_{av}^{2}B_{i,j}/ \bar{B})^{0.5}, i= 1,2,...,L,j= 1,2,...,K;
\end{equation}
where $\pmb{B}=[B_{i,j}]$ is a brightness of the  fragment of a randomly selected image from the training set, $\bar{B}$ is the mean of $\pmb{B}$.
The corresponding input patch $\pmb{P}$ is generated by: 
\begin{equation} \label{sigm3}
   P_{i,j} = \mathcal{N}(I_{i,j}, \sigma_{i,j}^{2}), i= 1,2,...,L,j= 1,2,...,K;
\end{equation}
where $\pmb{I}=[I_{i,j}]$ is an image fragment of size 128x128.

\begin{figure*}[!ht!]
 \centering
\setlength\tabcolsep{2pt}
\begin{tabular}{ccccccc}
 \centering
\includegraphics[width=0.27\columnwidth]{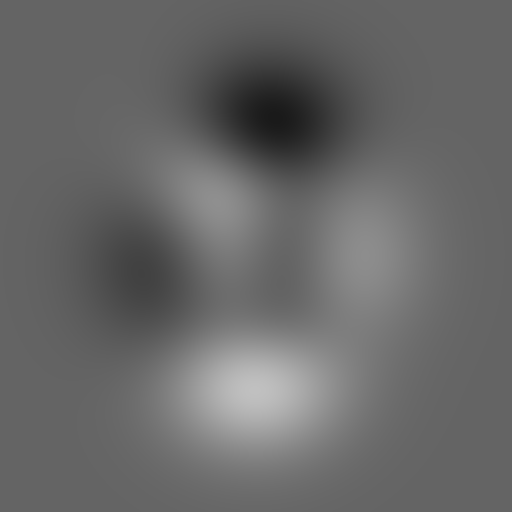} &
\includegraphics[width=0.27\columnwidth]{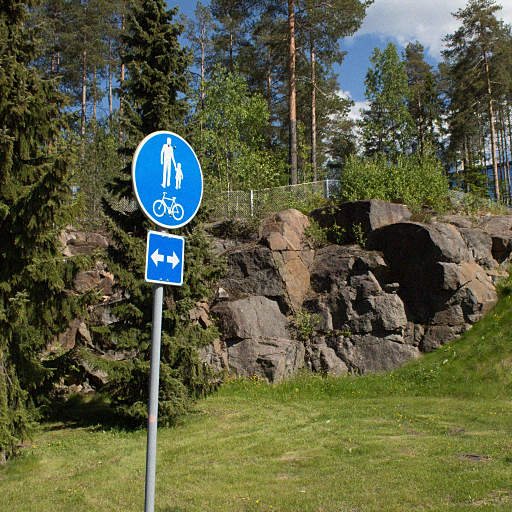} &
\includegraphics[width=0.27\columnwidth]{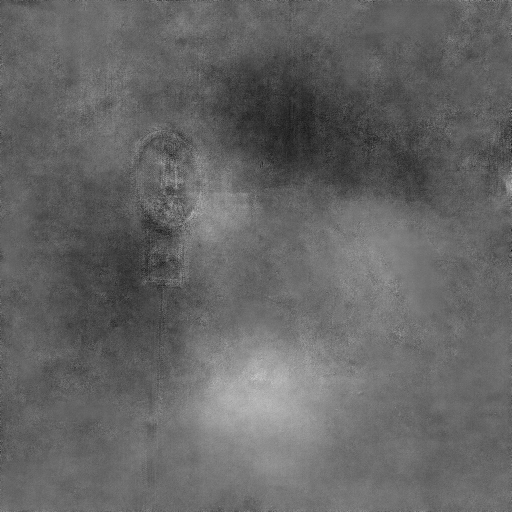} &
\includegraphics[width=0.27\columnwidth]{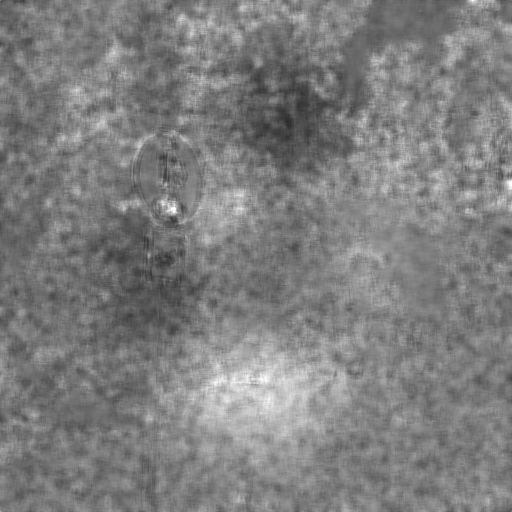} &
\includegraphics[width=0.27\columnwidth]{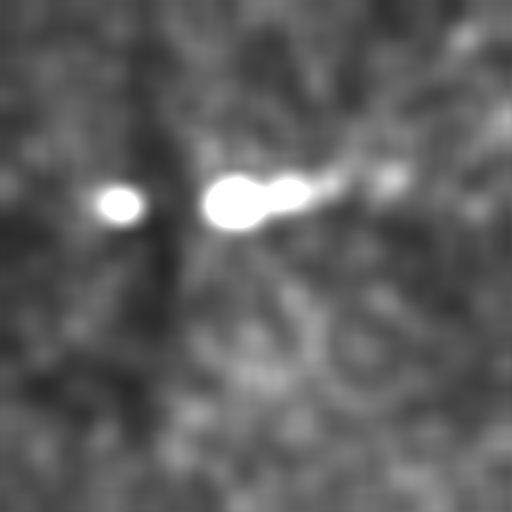} &
\includegraphics[width=0.27\columnwidth]{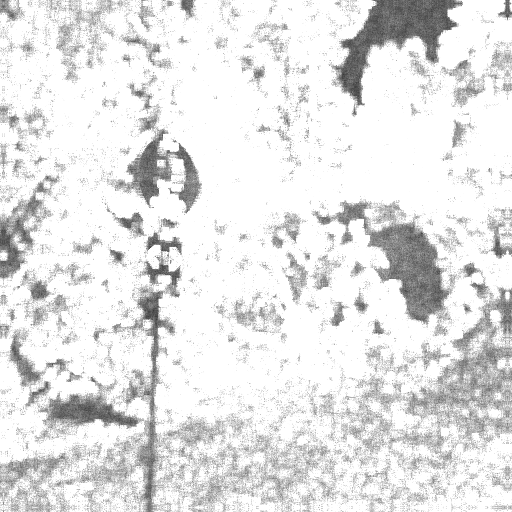} &
\includegraphics[width=0.27\columnwidth]{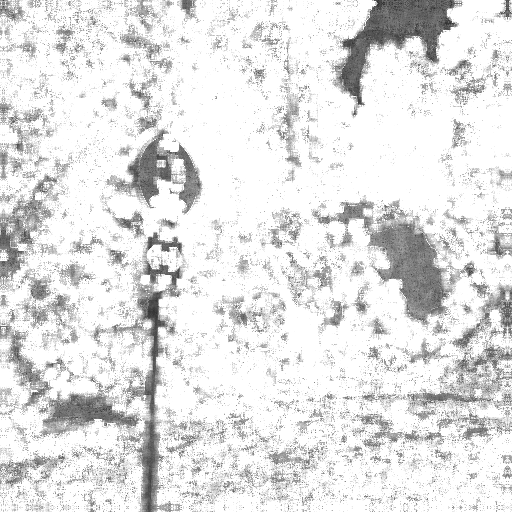} \\
 Ground truth & Noisy & SDNet & VDNet & DCAE & RHDCT & LADCT \\ 
 $\sigma_{av}$=5 & image & $\varepsilon_{m}$=0.12 & $\varepsilon_{m}$=0.27 & $\varepsilon_{m}$=0.44 & $\varepsilon_{m}$=2.0 & $\varepsilon_{m}$=2.2
\end{tabular} 
\caption{Ground truth sigma-map and the results of its estimation from the noisy image by different methods}
\label{Figure:maps}
\end{figure*}

Image fragments $\pmb{I}$ for generation of patches are  selected in the following way. First, a random image from the  training set is extracted. Next, a fragment of size 128x128 is cropped from this image 
to increase
the presence of patches with fine details and textures in the training set. This is done by the algorithm described in \cite{ponomarenko2020deep} (selecting three  image fragments). 
To decrease over-learning,  random rotations (on $90^{\circ}$, $180^{\circ}$, $270^{\circ}$) and mirroring of selected fragments are used.

The half of input patches $\pmb{P}$ are clipped by restricting pixel values to be in the [0...255] interval. This is done to make a pre-trained SDNet effective for both clipped and non-clipped noise cases. 

The design and training of SDNet has been carried out in Matlab R2021a, using the  custom training loop. 
 Overall, 150000 iterations with minibatch size 32 were performed. MSE has been used as a loss function to provide smaller number of outliers in sigma-map estimates.  Adam optimizer was used with the learning rate 
 $10^{-5}$ for the first 100000 iterations and $5*10^{-6}$  for the last 50000 iterations.
\subsection{Modification to color images}
SDNet has been separately trained for color and grayscale images.
To provide a compatibility with the setup used in \cite{yue2019variational}, 
for color images,   noise was independently added to each color component image according to (\ref{sigm3}) and by using the same sigma-map. 


\section{Numerical Analysis:Sigma-map Estimation}

In this section, we  analyze a quality of sigma-maps estimation by the  pre-trained SDNet. A set of 24 color images (22 images from Tampere17 dataset \cite{ponomarenko2018blind} in addition to Barbara and Baboon images) are selected for testing. Three non-stationary sigma-map models introduced in \cite{yue2019variational} are used to create a noisy test set. We have multiplied these sigma-maps by a factor to provide necessary mean variance $\sigma_{av}^{2}$ of a sigma-map. Thus, 72 noisy images are generated for each $\sigma_{av}$ value.

\subsection{Quality of estimation of sigma-map for non-stationary noise}

We have used $\varepsilon_{m}$ defined in (\ref{eq5}) as a  qualitative criterion ($n=72$). The relative errors are demonstrated in Table \ref{tab:tab2}, to  compare quality of sigma-map estimation by the proposed method SDNet, methods RHDCT \cite{shulev2006threshold},  LADCT \cite{lukin2010discrete}, DCAE \cite{bahncmiri2019deep}, and the state-of-the-art VDNet \cite{yue2019variational}.
 
In Table \ref{tab:tab2} we have also presented a grayscale modification of SDNet ('SDNet grayscale') for the comparison. Here sigma-maps are separately estimated for image color channels and then averaged. It is worth noting that estimation of noise for small \(\sigma_{m}\) is a difficult task. As one can see from the results provided by RHDCT and LADCT (see  example in  Fig.\ref{Figure:maps}), these methods do not provide satisfactory results, producing relative error values larger than 1. 

\begin{table*}[ht]
\caption{\label{tab:tab5} Image denoising in the case of  non-stationary noise, PSNR, dB and SSIM}
\setlength\tabcolsep{2pt}
\begin{tabular}{|c|c|c|c|c|c|c|c|c|c|c|c|c|c|c|c|c|c|c|c|c|}
\hline
 $\sigma_{av}$ &
  \multicolumn{2}{c|}{\begin{tabular}[c]{@{}c@{}}Noisy\\ images \end{tabular}} &
  \multicolumn{2}{c|}{\begin{tabular}[c]{@{}c@{}}CBDNet\\ (blind)   \end{tabular}} &
  \multicolumn{2}{c|}{\begin{tabular}[c]{@{}c@{}}DnCNN\\ (blind)  \end{tabular}} &
  \multicolumn{2}{c|}{\begin{tabular}[c]{@{}c@{}}  VDNet   \end{tabular}} &
  \multicolumn{2}{c|}{\begin{tabular}[c]{@{}c@{}}FFDNet+\\ color \\ SDNet   \end{tabular}} &
  \multicolumn{2}{c|}{\begin{tabular}[c]{@{}c@{}}FFDNet  + \\ true \\ sigma-map   \end{tabular}} &
  \multicolumn{2}{c|}{\begin{tabular}[c]{@{}c@{}}DRUNet+\\ VDNet \end{tabular}} &
  \multicolumn{2}{c|}{\begin{tabular}[c]{@{}c@{}}DRUNet+\\ grayscale \\ SDNet  \end{tabular}} &
  \multicolumn{2}{c|}{\begin{tabular}[c]{@{}c@{}}DRUNet+ \\ color \\ SDNet  \end{tabular}} &
  \multicolumn{2}{c|}{\begin{tabular}[c]{@{}c@{}}DRUNet+\\ true \\ sigma map \end{tabular}} \\ \hline
  & PSNR & SSIM  & PSNR & SSIM & PSNR & SSIM  & PSNR & SSIM & PSNR & SSIM  & PSNR & SSIM & PSNR & SSIM & PSNR & SSIM  & PSNR & SSIM& PSNR & SSIM \\ \hline
5 & 34.2 & 0.953 & 31.7 & 0.950 & 34.1 & 0.973 & 37.0 & 0.981 & 36.5 & 0.980 & 36.8 & 0.981 & 37.1 & 0.981 & \textbf{37.5} & 0.982 & \textbf{37.5} & \textbf{0.983} & 38.1 & 0.984 \\ \hline

7 & 31.3 & 0.919 & 31.5 & 0.947 & 33.3 & 0.966 & 35.3 & 0.973 & 35.0 & 0.971 & 35.2 & 0.973 & 35.5 & 0.974 & 35.6 & \textbf{0.975} & \textbf{35.8} & \textbf{0.975} & 36.1 &  0.977\\ \hline
10 & 28.3 & 0.864 & 31.0  & 0.938 & 32.1 & 0.954 & 33.5 & 0.961 &  33.2 & 0.959 & 33.3 & 0.959 & 33.7 &0.962 & 33.7 & \textbf{0.963} & \textbf{33.9} & \textbf{0.963} & 34.1 & 0.964 \\ \hline
15 & 24.8 & 0.774 &  29.8 & 0.917 & 30.5 & 0.933 & 31.4 & 0.941 & 31.1 & 0.936  & 31.2  & 0.936 & 31.6 & 0.940 & 31.6 & 0.941 & \textbf{31.7} & \textbf{0.942} & 31.8 & 0.943 \\ \hline
20 & 22.4 & 0.693 & 28.6  & 0.893 & 29.3 & 0.912 & 29.9 & \textbf{0.920} & 29.6 & 0.912 & 29.7 & 0.912 & 30.0 & 0.918 & \textbf{30.1} & 0.919 & \textbf{30.1} & \textbf{0.920}  & 30.2 & 0.920 \\ \hline
30 & 19.1 & 0.562 & 26.8  & 0.847 & 27.4  & 0.870 & 27.8 & \textbf{0.880} & 27.5 & 0.863 & 27.5 & 0.864 & \textbf{27.9} & 0.876 & \textbf{27.9} & 0.875 & 27.8  & 0.876 & 27.9 & 0.876 \\ \hline
45 & 16.0 & 0.424 &  25.0 &  0.783 & 25.3  & 0.813 & 25.6 & \textbf{0.826} & 25.1 & 0.793 & 25.1 & 0.794 & \textbf{25.7} & 0.820 & 25.5 & 0.811 & 25.4  & 0.812 & 25.5 & 0.813
 \\ \hline
\end{tabular}
\end{table*}

\begin{table}[ht]
\caption{\label{tab:tab2} Relative estimation error $\varepsilon_{m}$ of sigma-maps}
\centering
\begin{tabular}{|c|c|c|c|c|c|c|}
\hline
$\sigma_{av}$ & LADCT & RHDCT & DCAE & VDNet & \begin{tabular}{@{}c@{}}SDNet \\ grayscale\end{tabular} & \begin{tabular}{@{}c@{}}SDNet \\ color\end{tabular}\\
\hline
5 & 1.87 & 1.69 & 0.41 & 0.31 & 0.27 & \textbf{0.21} \\
\hline
7 & 1.24 & 1.11 & 0.32 & 0.21 & 0.19 & \textbf{0.13} \\
\hline
10 & 0.80 & 0.69 & 0.28 & 0.15 & 0.14 & \textbf{0.08} \\
\hline
15 & 0.49 & 0.40 & 0.22 & 0.11 & 0.08 & \textbf{0.05} \\
\hline
20 & 0.35 & 0.27 & 0.21 & 0.10 & 0.06 & \textbf{0.04} \\
\hline
30 & 0.24 & 0.19 & 0.25 & 0.11 & 0.04 & \textbf{0.03} \\
\hline
45 & 0.20 & 0.19 & 0.31 & 0.13 & \textbf{0.03} & \textbf{0.03} \\
\hline
\end{tabular}
\end{table}

\subsection{Quality of estimation of standard deviation of AWGN}

The proposed SDNet can be applied for a  special case when all $\sigma_{i,j} = \sigma$, which corresponds to the case of i.i.d. Gaussian noise model. 

For a comparative amalysis we have  selected three methods: IEDD \cite{ponomarenko2018blind}, PCA \cite{pyatykh2012image} and WTP \cite{tanaka2012}, which are state-of-the-art for estimation of standard deviation (STD) of AWGN. We have also included in this analysis VDNet \cite{yue2019variational}.
Estimated STD for a given image by SDNet and VDNet methods are calculated as medians of the  corresponding estimated sigma-maps. 

As a quality criterion we used a relative error of estimation of STD, defined by:
\begin{equation} \label{eq6}
        \varepsilon  =  \frac{ ||\pmb{\sigma_{e}}-\sigma_{t}||_{2}}{n\sigma_{t}} 
\end{equation}
where $\sigma_{t}$ is a  true value of AWGN STD, $n$ is a number of estimates (here $n=24$), $\pmb{\sigma_{e}}$ is a vector of estimated standard deviations for test images. Both clipped and non-clipped noise are considered in this analysis (see Tables \ref{tab:tab3} and \ref{tab:tab4}). 

SDNet attains the best accuracy, outperforming other methods almost for all cases. Moreover, for $\sigma_{t}$ equal to 3 and 5, SDNet estimation error is twice smaller than that of the nearest competitor. 
It is interesting to observe that all four methods IEDD, PCA, WTP and VDNet fail for large $\sigma_{t}$ for clipped noise case (values are marked in Table \ref{tab:tab4} in   italic and  underlined), while SDNet demonstrates very  good accuracy both for clipped and non-clipped noise cases.


\section{Numerical Analysis: Denoising Using Estimated Sigma-map}

Here we use  estimated  sigma-maps in image denoising. Comparing  denoised images obtained by the  different sigma-map estimators, we  show how much one can boost the  performance of denoising by improving accuracy of sigma-map estimation. 
As denoisers in this study we use  DRUNet and FFDNet\cite{zhang2018ffdnet},
since their network architecture 
accepts noise sigma-map as an auxiliary input. Three state-of-the-art CNN-based blind denoising methods, DnCNN\cite{zhang2017beyond}, CBDNet\cite{guo2019toward} and VDNet, have been selected for the 
comparison. The peak signal-to-noise ratio (PSNR) and structural similarity index measure (SSIM) are used as quality criteria.

Note that DnCNN was originally designed to address the problem of Gaussian denoising with unknown noise level, while CBDNet is taking one step further and includes  real-world noisy-clean image pairs as training input to estimate more realistic noise models. VDNet is also trained to handle a  non-stationary noise, it  can simultaneously estimate sigma-map and perform blind denoising. Table \ref{tab:tab5} shows average PSNR and SSIM values for 72 denoised test images. 
Furthermore, for denoisers equipped with the sigma-map estimation by the proposed SDNet, we provide also the results for ideal case of the denoising when the true sigma-map are used. 
Small differences between PSNR and SSIM values of the results produced by the denoisers accepting true sigma-map with the same denoisers accepting SDNet estimated sigma-map,  
demonstrate the high accuracy of sigma-map estimation and its influence on denoising performance.  
As one can see from Table \ref{tab:tab5}, a  combination of SDNet with DRUNet yields the best denoising performance,and the gap between best results and the results from ideal case (using true sigma-maps) is less than 0.6 db in PSNR and none in SSIM.
Although the same comparison is performed for the case of FFDNet denoiser and the gap is smaller, quality of denoising using FFDNet is about 1 db lower than that of DRUNet. In the case of blind denoisers, quality of denoised images is much lower, within the range of 2 $\sim$  6 dB for low noise levels. 
It is also interesting that for small noise levels DRUNet with sigma map estimated by VDNet provides better results then VDNet denoiser trained jointly with the sigma-map estimation network. 
One can observe from the denoised images that blind denoisers can 
distort texture and eliminating fine details of an image. Denoising based on sigma-map estimated by the proposed SDNet overcomes this problem and preserves fine image details in the low noise levels.

\begin{table}[ht]
\caption{\label{tab:tab3} $\varepsilon$ for non-clipped AWGN}
\centering
\begin{tabular}{|c|c|c|c|c|c|}
\hline
true $\sigma$ & IEDD & PCA & WTP & VDNet & SDNet \\
\hline
3 & 0.407 & 0.573 & 0.508 & 1.783 & \textbf{0.200} \\
\hline
5 & 0.204 & 0.272 & 0.271 & 0.906 & \textbf{0.106} \\
\hline
7 & 0.110 & 0.167 & 0.176 & 0.560 & \textbf{0.073} \\
\hline
10 & 0.054 & 0.098 & 0.126 & 0.325 & \textbf{0.043} \\
\hline
15 & 0.033 & 0.057 & 0.079 & 0.171 & \textbf{0.028} \\
\hline
20 & \textbf{0.021} & 0.044 & 0.057 & 0.110 & 0.029 \\
\hline
30 & \textbf{0.014} & 0.031 & 0.029 & 0.063 & 0.018 \\
\hline
50 & 0.013 & 0.022 & 0.012 & 0.040 & \textbf{0.011} \\
\hline
75 & 0.013 & 0.023 & \textbf{0.007} & 0.032 & \textbf{0.007} \\
\hline
\end{tabular}
\end{table}

\begin{table}[ht]
\caption{\label{tab:tab4} $\varepsilon$ for clipped AWGN}
\centering
\begin{tabular}{|c|c|c|c|c|c|}
\hline
true $\sigma$ & IEDD & PCA & WTP & VDNet & SDNet \\
\hline
3 & 0.407 & 0.573 & 0.507 & 1.782 & \textbf{0.200} \\
\hline
5 & 0.204 & 0.274 & 0.269 & 0.905 & \textbf{0.110} \\
\hline
7 & 0.114 & 0.166 & 0.174 & 0.559 & \textbf{0.076} \\
\hline
10 & 0.073 & 0.097 & 0.122 & 0.323 & \textbf{0.043} \\
\hline
15 & 0.071 & 0.053 & 0.075 & 0.168 & \textbf{0.026} \\
\hline
20 & \underline{\textit{0.077}} & 0.038 & 0.057 & 0.105 & \textbf{0.027} \\
\hline
30 & \underline{\textit{0.114}} & 0.036 & 0.055 & 0.066 & \textbf{0.017} \\
\hline
50 & \underline{\textit{0.157}} & \underline{\textit{0.069}} & \underline{\textit{0.103}} & \underline{\textit{0.089}} & \textbf{0.014} \\
\hline
75 & \underline{\textit{0.207}} & \underline{\textit{0.172}} & \underline{\textit{0.176}} & \underline{\textit{0.150}} & \textbf{0.010} \\
\hline
\end{tabular}
\end{table}

\section{Conclusion}	
We propose an effective CNN-based method, SDNet,  for sigma-map  estimation. Advantages of this CNN architecture were  discussed. The training set generation and selection of training patches were described. \\
We have compared the proposed method with the state-of-the-art methods. The results show that 
SDNet outperforms them in   estimation accuracy for both sigma-maps and standard deviation of AWGN, and for both   clipped and non-clipped noise cases. For small levels of noise estimation errors for SDNet in average are twice smaller than those of nearest competitors.

A comparative analysis of denoising efficiency using estimated sigma maps demonstrates that the combination of SDNet and DRUNet provides the  state-of-the-art quality of image  denoising which is very close to the ideal case (where true sigma-maps are used).

\vfill


\bibliographystyle{IEEEtran}
\bibliography{IEEEabrv,Bibliography}

\end{document}